\documentclass{ws-ijmpb}
\begin{document}

\markboth{Leticia F. Cugliandolo}{Dissipative quantum disordered models}

\catchline{}{}{}{}{}

\title{Dissipative quantum disordered models}

\title{Dissipative quantum disordered models\footnote{Contribution to 
the 12th International
Conference on Recent Progress in Many-Body Theories. Santa Fe, New Mexico, 
23-27 August 2004}}
\author{Leticia F. Cugliandolo}
\address{Laboratoire de Physique Th\'eorique et Hautes Energies, Jussieu, 
France and 
\\
Laboratoire de Physique Th\'eorique, Ecole Normale 
Sup\'erieure, Paris, France
\\
leticia@lpt.ens.fr}
\maketitle 
\begin{history}
\received{13 September 2004}
\end{history}

\begin{abstract}
This article reviews recent studies of mean-field and one dimensional
quantum disordered spin systems coupled to different types of
dissipative environments. The main issues discussed are: (i) The
real-time dynamics in the glassy phase and how they compare to the
behaviour of the same models in their classical limit.  (ii) The phase
transition separating the ordered -- glassy -- phase from the
disordered phase that, for some long-range
interactions, is of second order at high temperatures and of first
order close to the quantum critical point (similarly to what has been
observed in random dipolar magnets). (iii)
The static properties of the Griffiths phase in random Ising
chains. (iv) The dependence of all these properties on
the environment.  The analytic and numeric techniques
used to derive these results are briefly mentioned.
\end{abstract}



\section{Introduction}
\label{intro}

Glasses slowly evolve towards equilibrium though never reaching it in
observable times scales.  Scientific research in this area started
more than a century ago. The amount of experimental data gathered is
huge.  Systems that would reach equilibrium in observable time-scales,
such as weakly sheared complex liquids, can be driven out of
equilibrium by external perturbations and still evolve slowl.y Powders
stay in static metastable states unless externally tapped or sheared:
these non-equilibrium perturbations slowly drive them  towards more compact
configurations.  Even if {\it a priori} very different, these systems
share many dynamic properties.

In several cases of practical interest quantum effects play an
important role.  On the experimental side spin-glass phases have been
identified in many condensed matter systems at very low
temperature. Among them we can cite the bi-layer Kagome
system\cite{Kagome} S$_r$Cr$_s$Ga$_4$O$_{19}$, the polychlore
structure\cite{pyrochlore} Li$_x$Zn$_{1-x}$V$_2$O$_4$, the dipolar
magnet\cite{aeppli,aeppli-first} LiHo$_x$Y$_{1-x}$F$_4$ and the high
$T_c$ compound\cite{highTc} La$_{1-x}$Sr$_2$Cu$_2$O$_4$. Quantum
glassy phases exit also in electronic systems\cite{Zvi} 
and structural glasses\cite{Osheroff} such
as Mylar.  The driven case is also very
important in quantum systems, think of an electronic device driven by
an external current.

The non-equilibrium dynamics of classical glasses, 
weakly driven complex liquids
and granular matter have been rationalized within a theoretical
approach that is based on the solution of mean-field simple 
models.\cite{LesHouches}
How much of the classical glassy phenomenology survives at very low
temperatures where quantum effects are important is a question that
deserves careful theoretical and experimental analysis.  The
impossibility of simulating the real-time evolution of quantum systems
of moderate size enhances the importance of {\it solving} simple
mean-field or low dimensional models.

On the other hand, peculiar phenomena in quantum phase transitions
have been signaled analytically and experimentally in systems with and
without quenched disorder.\cite{Sachdev-book,Senthil}  For instance,
at low temperatures and intermediate dipole concentration, the
dipolar-coupled Ising magnet LiHo$_x$Y$_{1-x}$F$_4$ in a transverse
field exhibits a spin-glass-like phase.\cite{aeppli,aeppli-first} The
phase transition  is of second order at low transverse field but
becomes first order close to the quantum critical
point.\cite{aeppli-first}  Morever, the relaxation in the glassy phase
is extremely slow and has very strong memory effects.

Another hallmark of 
finite dimensional disordered quantum spin models are
Griffiths-McCoy
singularities, that lead to a  highly non trivial paramagnetic phase
and critical behaviour.

In this contribution we summarize the results of 
recent studies of
the statics, dynamics and critical properties
of mean-field and one dimensional quantum
disordered spin systems coupled to an environment.\cite{Culo-Culolo}
We also briefly mention related studies on a driven mesoscopic
ring,\cite{Lili} dilute antiferromagnets,\cite{Malcolm} and a manifold
in an infinite dimensional quenched random potential.\cite{Pierre} In
Sect.~\ref{questions} we review the main questions that we attempted
to answer in these papers. In Sect.~\ref{models} we recall the
definition of the models that we studied. 
The basic techniques used to study classical glassy models with or
without disorder are well documented in the literature (the replica
trick, scaling arguments and droplet theories, the dynamic functional
method used to derive macroscopic equations from the microscopic
Langevin dynamics, functional renormalization, Montecarlo and
molecular dynamic numerical methods). On the contrary, the techniques
needed to deal with the statics and dynamics of quantum macroscopic
systems are much less known in general.  We briefly
mention the ones that we used to study 
dissipative disordered quantum models.\cite{Boulder}
Finally, in Sect.~\ref{perspectives} we list some
projects for future research.

\section{Questions and main results}
\label{questions}


\subsection{Effect of quantum fluctuations on glassiness}

Since glasses are not expected to reach equilibrium in experimentally
accessible times, it is important to device a method to understand the
influence of quantum fluctuations on their trully nonequilibrium {\it
real-time} dynamics.  Intuitively, one expects quantum fluctuations to
only affect the short-time dynamics; however, they are also expected
to act as thermal fluctuations. It is then not clear {\it a priori}
whether quantum fluctuations would tend to destroy glassiness or
modify it drastically.

The usual methods of equilibrium quantum statistical mechanics are
inappropriate to describe this nonstationary situation. 
We presented a formalism suited to study the real-time dynamics of a
general nonlinear, possibly disordered, model in contact with a
bath that can also be applied to glassy models.\cite{Culo} 
The method is a combination of the Schwinger-Keldysh
or closed-time path technique to study real-time phenomena, with the
Feynman-Vernon approach to dissipation that consists in modelling the
coupling to the environment with an ensemble of quantum harmonic
oscillators.  As a particular case we studied the relaxation of the
spherical version of the quantum $p$ spin fully-connected disordered
model. We analyzed the relaxation of `random initial
conditions' in the limit of vanishing coupling strength taken after
the long waiting-time ($t_w$) limit.  The same technique was applied to
other quantum problems\cite{Lev}$^-$\cite{Premi}
and related studies appeared.\cite{other-quantum}$^-$\cite{Galitsky}
We later studied the effect of a strong coupling to the
environment\cite{bath-spherical} ({\it i.e.} $t_w\to\infty$
with $\alpha$ finite) as discussed below.

In the disordered phase the dynamics is fast and occurs in
equilibrium.  The correlation and linear response are stationary,
{\it i.e.} $C(t + t_w,t_w)=C(t)$
and $R(t+t_w,t_w)=R(t)$. They
both oscillate with a $\Gamma$-dependent 
frequency that also depends on 
the characteristics of the bath $(\alpha,s)$
if $\alpha$ is not taken to zero.  
Correlations and responses are linked by the quantum
fluctuation dissipation theorem ({\sc fdt}).  At high temperatures and
after a short transient the system decoheres and the dynamics becomes
classical ({\it e.g.} responses and correlations are related
by the classical {\sc fdt}).

In the ordered phase the glassy dynamics persists asymptotically if
the thermodynamic limit has been taken at the outset.  The dynamics of
glassy systems occurs out of equilibrium and the correlations and
responses loose time translation invariance.  If $t_w$ denotes the
time elapsed since a quench into the {\sc sg} phase, $C(t + t_w,t_w)$
and $R(t+t_w,t_w)$ depend on both $t$ and $t_w$. The order in which
the limits $t_w\to\infty$ and $t\to\infty$ are taken is very
important. For sufficiently long $t$ and $t_w$ but in the regime $t
\ll t_w$, the dynamics is stationary and the correlation reaches a
plateau $q_{\sc ea}$.  A few quantum oscillations exist at very short
$t$ (and arbitrarily large value of $t_w$) and they later disappear.
For times $t > t_w$, the system enters an {\it aging} regime where the
correlation function depends on $t_w$ explicitly.  In
this regime, the correlation vanishes at long times,
$\lim_{t \to \infty} C(t + t_w,t_w) = 0$, at a rate that depends on
$t_w$. The behaviour is thus qualitative similar to what observed
classically, even if the scaling laws are modified by the quantum
fluctuations. One checks that the relaxation of typical 
(highly energetical) initial conditions approaches a threshold 
level in phase space with a higher energy-density than the equilibrium 
one.

The comparison of responses and correlations in the ordered phase is
particularly interesting. The quantum {\sc fdt} is a complicated
integral relation between the correlation and the linear
response. This relation holds for $t\ll t_w$ when the correlation
decays to the plateau. In the second regime the quantum {\sc fdt} is
no longer verified, much as it happens in the classical problem. A
comparison of the integrated responses and the symmetrized correlation
in a parametric manner\cite{LesHouches} shows that the two quantities
are related by a {\it classical} {\sc fdt} with an effective
temperature\cite{Cukupe} $T_{\sc eff}$ that depends on the parameters
in the problem $(T, \Gamma)$ [and the characteristics of the bath,
$\alpha, s$].  We proved that $T_{\sc eff}>T$ and it is different from
zero even when the environment is at $T=0$.  $T_{eff}$ drives the
dynamics at late epochs and it makes the dynamics appear classical in
that two-time regime.  $T_{eff} > 0$ even
when the temperature of the bath is zero.  The generation of a
non-trivial $T_{eff}$ for the slow part of the decay gives
support to the ``decoherent'' effect observed in the decay of
correlations.

The similarity between the second decay in the classical and quantum
problem can be argued as follows. The responses decay rather
fast to zero when the time difference increases (though
integrated over a time-interval of very long length does not
vanish.\cite{Cuku}) The explicit dependence on $\hbar$ in the regime
of widely separated times comes from factors with higher powers of $R$
that vanish\cite{Culo}.  The effect of quantum fluctuations on the
slow time-difference regime is simply to renormalize\cite{Kech}
certain parameters in the equations that otherwise 
look classical.

Models with $p\geq 3$ interactions have different static and dynamic
phase transitions, with the latter surrounding a larger region of
the $(T, \Gamma)$ phase diagram.  The static and dynamic
ordered-disorder phase transition present a second-to-first order
transition.\cite{Niri,Cugrsa1} Close to the classical critical
temperature quantum effects are small and the phase transition is
discontinuous but of second order, as in the classical case.  There
are no discontinuities in the thermodynamic quantities but there is a
plateau that develops in the correlation function when the transition
is approached from the disordered side. This is the behaviour 
expected 
in classical glasses. Spin-glasses instead have continuous transitions
(without precursors).  Conversely, close to the quantum critical point
quantum fluctuations drive the transition first order
thermodynamically. Across the first order line the susceptibility is
discontinuous and shows hysteresis.  This is similar to what has been
observed in the dipolar-coupled Ising magnet LiHo$_x$Y$_{1-x}$F$_4$ in
a transverse field.\cite{aeppli-first}

We also adapted the {\it Ansatz} of marginal 
stability\cite{Cugrsa1,ams,Niri} to
identify the dynamic critical line that is consistent with the one
found using the Schwinger-Keldysh formalism. 
The analytic continuation of the imaginary-time dependent
correlation computed with the {\sc ams} in the absence of
the bath is identical to the {\it stationary} part of the
non-equilibrium correlation function ($C>q_{\sc ea}$) when one takes
the long-time limit first and the limit in which the coupling to the
bath goes to zero next.\cite{Cugrsa1,Pierre}  

In the classical case, the study of the Thouless-Anderson-Palmer ({\sc
tap}) free energy landscape has been very useful to understand the
behaviour of these systems.\cite{LesHouches} A {\sc tap} approach can
also be developed for quantum problems.\cite{Bicu} It helps
understanding the existence of a dynamic and a static critical line as
well as the change in nature of the transition close to the quantum
critical point.

\subsection{Effect of the bath: decoherence and localization}

The quantum systems mentioned in Sect.~\ref{intro}
are not totally isolated but in
contact with environments of different type.

The low-energy physics of many tunneling systems is well described by
the spin-boson model in which the two equivalent degenerate states are
represented by the eigenstates $\sigma_z = \pm 1$ of an Ising
pseudo-spin. A transverse field coupled to $\sigma_x$ (say) represents
the tunneling matrix element. The coupling to the environment is given
in terms of its spectral density $I(\omega)\propto \alpha\,\omega^s$
for $\omega \ll \omega_c$, where $\alpha$ is a dimensionless coupling
constant and $\omega_c$ a high frequency cutoff.  The exponent $s$
characterizes different types of environment.  The Ohmic case ($s =
1)$ is quite generally encountered but superOhmic ($s>1$) and subOhmic
($s<1$) baths also occur in, {\it e.g.}, the Kondo effect
in unconventional hosts.

The coupling of quantum two-level systems ({\sc tls}) to a dissipative
environment has decisive effects on their dynamical properties. The
dilute case, in which interactions between the {\sc tls} can be
neglected, has been extensively
investigated.\cite{Standard,Leggett-review} This problem, is related
the $1d$ Ising model with inverse squared interactions and the
anisotropic Kondo model.  In the Ohmic case, at zero temperature,
there is a phase transition\cite{Bray-Moore} at $\alpha=1$.  For
$\alpha<1$ there is tunneling and two distinct regimes develop. If
$\alpha<1/2$ the system relaxes with damped coherent oscillations; in
the intermediate region $1/2<\alpha<1$ the system relaxes
incoherently.  For $\alpha>1$ quantum tunneling is suppressed and
$\langle \hat \sigma_z\rangle \neq 0$ signalling that the system
remains localized in the state in which it was prepared.  These
results also hold for sub-Ohmic baths while weakly damped oscillations
persist for super-Ohmic baths.  At finite temperatures (but low enough
such that thermal activation can be neglected), there is no
localization but the probability of finding the system in the state it
was prepared decreases slowly with time for $\alpha>\alpha^{\sc
c}$.

The effect of dissipation on the phase transition, critical behaviour,
ordered phase, localization and decoherence properties of macrocopic
interacting systems is only now starting to be
analyzed.\cite{bath-spherical,bath-SU2,Troyer}
In thermodynamic equilibrium, in the absence of the bath, the
interactions between the {\sc tls} lead to the appearance of an
ordered state at low enough temperature. If the
interactions are of random sign, as in the models we considered,
the latter will be a glassy state. In this phase
the symmetry between the states $\sigma_i^z = \pm 1$
 at any particular site is broken but there is no global
magnetization, $\sum_i \langle \hat \sigma_i^z \rangle = 0$.
The coupling to the bath also 
competes with the tunneling term. 
We thus expect the presence of noise to increase the
stability of the glassy state. The
consequences of this fact are
particularly interesting when there is 
localization at some $\alpha=\alpha^{\sc c}$:
a quantum critical point at ${J}=0$,
$\alpha=\alpha^{\sc c}$ separates the disordered and the ordered
state such that, for  $\alpha > \alpha^{\sc c}$, the glassy 
phase survives down to ${J}=0$.

A system of non-interacting localized {\sc tls} and a glassy state
{\it in equilibrium} are in some way similar
However, this resemblance is only superficial.
The details of the dynamics of the two systems are expected to be
quite different, with $C$ saturating at a finite value in the localized 
state, and $C$ decaying down to zero in the glassy phase.

\subsection{Effect of the bath on interacting mean-field 
models} 
 
The problem of a single {\sc tls} being a difficult one, that of an
infinite set of interacting {\sc tls} seems hardly solvable.
Therefore, as a first step, we focused on the effect of the reservoir
on the $p$-spin {\it spherical} model,\cite{bath-spherical} a problem
that we studied with the real-time approach and the replica Matsubara
technique. The position of the critical lines strongly depends on the
strength of the coupling to the bath and the type of bath (Ohmic,
subOhmic, superOhmic).  For a given type of bath, the ordered glassy
phase is favored by a stronger coupling.  The classical static and
dynamic critical temperatures 
remain unchanged by the coupling to the environment. The identity 
between the analytic continuation  of the 
imaginary-time correlation to real-time and the correlation 
in the Schwinger-Keldysh approach also holds if the strength of the 
bath is finite.

The spherical model localizes in the absence of interactions when
coupled to a subOhmic bath: $C(t+t_w,t_w)$ reaches, for any
waiting-time $t_w$ and long enough $t$, a plateau that it never
leaves.  When interactions are switched on localization disappears and
the system undergoes a phase transition towards a glassy phase.

Similar results were found for the $SU(N)$ random Heisenberg model in
the limit $N\to\infty$\cite{Bipa}, the $p=2$ spherical
model\cite{Premi} and the SU(2) $p$-spin model.\cite{bath-SU2}

\subsection{Effect of disorder: Griffiths singularities}

Griffiths singularities\cite{Griffiths}$^-$\cite{McCoy} 
in classical finite dimensional random systems
are so weak that their consequences have not been clearly observed neither
experimentally nor numerically. Instead, when quantum
fluctuations are introduced they are much stronger 
and rare regions completely determine the static and
dynamic behaviour in the Griffiths phase.

The isolated quantum random Ising chain  has been 
studied in great detail with a decimation technique.~\cite{dfisher}
It undergoes a quantum phase transition from a {\sc pm} to a ferromagnet
for a special relation between the
distribution of exchanges and transverse fields. The quantum phase
transition is of second order with 
the correlation time scaling exponentially with the
spatial correlation length ({\it activated} scaling)
Within the renormalization group procedure this is a characteristic of 
an ``infinitely strong disorder'' fixed point.\cite{dfisher}
Within the {\sc pm} Griffiths phase 
the distributions of local linear and non-linear
magnetic susceptibilities are large and typical and average values
are very different, with the latter being dominated by rare regions.
The behaviour is higher spatial dimensions is similar.

The analysis of Montecarlo simulations of the equivalent $d+1$ 
classical Ising model is quite tricky. Initially, 
it was claimed that there was conventional scaling in
$d=1$\cite{1dMC} as well as in $d>1$\cite{2dsim}
but  a more careful analysis of the numerical data confirmed the activated
scaling in both cases.\cite{1dfermions}

On the real-time dynamic side, there have been some studies of the
relaxation of special initial conditions of the isolated random Ising
chain at constant energy.\cite{IgloiRieger}

\subsection{Disorder and dissipation: fate of Griffiths phase?}

If the interactions between two-level systems placed in a finite
dimensional space are random one may wonder what is the effect of the
bath on the Griffiths phase. The answer to this question has been
debated over the last years.\cite{Antonio-etal}$^-$\cite{Miranda}
We addressed this problem using Montecarlo
simulations of the equivalent $2d$ classical system.~\cite{Culolo} 
Preliminary results for rather small systems ($N_x\leq 32$, $N_\tau\leq 256$)
show that an Ohmic bath favors the glassy phase. 
Our results are compatible with (but we do not prove) 
activated scaling at criticality, 
at least for small values of $\alpha$. 

\section{The models}
\label{models}

Disordered quantum spin-$\frac12$ models with two-body 
interactions are defined by 
\begin{equation}
H_S = - \sum_{ij} J_{ij} \hat \sigma^z_i \hat \sigma^z_j
+ \sum_i \Gamma_i \hat \sigma_i^x + \sum_i h_i \hat \sigma_i^z
\; .
\label{HS}
\end{equation}
$i=1,\dots, N$ labels the spins that lie on the
vertices of a cubic $d$ dimensional lattice
and are represented by Pauli matrices. The interaction strengths
$J_{ij}$ couple near-neighbours only and are chosen from a probability
distribution, $P(J)$. The average and variance are defined as
$[J_{ij}]=J_o$ and $[J_{ij}^2]=J^2/(2c)$, where $J_o$ and $J$
are $O(1)$ and $c=2d$ is the connectivity of the lattice.  The
next-to-last term is a coupling to a random quenched local transverse field
$\Gamma_i$. The last term is the coupling to a longitudinal field
that serves to compute local susceptibilities.

Several generalizations that render the model easier to treat
analytically are:

\noindent
-- {\it Fully-connected limit.}  One allows each spin to interact with
all others, $c\to N-1$.

\noindent
-- {\it Multi-spin interactions.}
In the fully-connected case one can considers 
\begin{equation}
H_S = - \sum_{i_1\dots i_p} J_{i_1\dots i_p} 
\hat \sigma^z_{i_1} \dots \hat \sigma^z_{i_p} 
+ \sum_i \Gamma_i \hat \sigma_i^x 
+ \sum_i h_i \hat \sigma_i^z
\; .
\end{equation} 
where the sum runs over all
$p$-uplets,\cite{pspin,Niri,Cugrsa1,bath-SU2} with $p$ an integer
parameter, $p \geq 2$.  The exchanges are random independent variables
with variance $p!{J}^2/(2 N^{p-1})$.  This model provides a mean-field
description of the structural glass transition and glassy physics that
is also intimately related to the mode-coupling
approach.\cite{LesHouches} In its dilute limit, with no geometry 
but finite connectivity on each site, this model is related
to the $K$-sat optimization problem.~\cite{Zecchina}

\noindent
-- {\it Spherical variables -- a particle in a random 
potential.}
One considers the spherical limit, 
$\sum_{i=1}^N \langle \hat \sigma_i^2\rangle =N$, in which 
the $\hat \sigma_i$ may be interpreted as the coordinates of a particle 
moving on an
$N$-dimensional sphere. A kinetic term,
$K = \sum_{i=1}^N \hat P_i^2/(2M)$, is then 
included in the Hamiltonian, 
with $\hat P_i$ the conjugated momentum satisfying the commutation rules
$[\hat P_i, \hat P_j] = 0$, 
$[\hat P_i, \hat \sigma_j] = - i \hbar \delta_{ij}  
$.
Other spherical models have been discussed.\cite{Theo-spherical}

The coupling to the environment is modelled by
$H = H_S + H_B + H_I + H_{CT}$,
where $H_B$ is the Hamiltonian of the bath, $H_I$ represents the
interaction between the system and the bath and 
$H_{CT}=\sum_{l=1}^{\tilde{N}}(2 m_l \omega_{l}^2)^{-1} 
(\sum_{i=1}^{N} c_{i l} \hat{\sigma}_i^z )^2$ is a
counter-term that is usually added to eliminate an undesired mass
renormalization induced by the coupling.\cite{Standard}  
We assume that each spin
is coupled to its own set of $\tilde{N}/N$ independent harmonic
oscillators with $\tilde N$ the total number of them.  
For simplicity we consider the
 bilinear coupling,
$H_{I} = -\sum_{i=1}^N  \hat{\sigma}_i^{z} \sum_{l=1}^{\tilde{N}} 
c_{i l} \hat{x}_{l} $.
 For $p=2$ the fully-connected limit with two-body interactions 
models metallic spin-glasses.\cite{Kondo} 

\section{Perspectives}
\label{perspectives}

In this article we reviewed recent studies of insulating disordered magnets.

The principal merit of the fully-connected models is that they are
simple enough to be studied in detail. Yet, many of their properties
are generic and expected to hold at least qualitatively for more
realistic cases.\cite{LesHouches} The analysis of the fully-connected
models is by now quite complete, having applied the replica theory,
the real-time dynamics approach, and the investigation
of the {\sc tap} free-energy landscape. A problem that remains 
not fully developed though is the treatment of the relaxation of initial 
conditions that are correlated with disorder.~\cite{Cugrsa2}

Recently, much progress has been done in the study of the statics of
classical dilute spin models, that is to say, models defined on random
hyper graphs with finite connectivity. The statics of these models 
encode problems in combinatorial opimization such
as $K$-sat.\cite{Zecchina} An interesting mapping relates (isolated)
dilute quantum disordered spin systems to the dynamics of
special purpose algorithms used in combinatorial optimization -- such
as Walk-Sat.\cite{walk-sat} It would be interesting to study the
latter using tools developed for the former and {\it vise versa}.

The great challenge remains to understand the behaviour of glassy
systems with and without disorder in finite dimensions. In particular,
one could try to adapt the decimation 
technique\cite{dfisher} to study disordered spin$-\frac12$ models
in contact with an environment.
One can also envisage applications of the ideas described in this
paper to other quantum systems evolving out of equilibrium.  In this
respect, we have studied the generation of an effective temperature in
a conducting ring threaded by time-dependent magnetic field and
coupled to a reservoir;\cite{Lili} and we plan to analyse quantum dilute
antiferromagnets\cite{Malcolm} and the low-temperature
dynamics of the Bragg glass,\cite{Pierre} as well as other related physical 
problems.

\section*{Acknowledgements}

This contribution summarizes work
done in collaboration with G. Biroli,
D. Grempel, G. Lozano, H. Lozza and C. da Silva Santos. I
also wish to thank L. Arrachea, C. Chamon, T. Giamarchi, L. Ioffe,
M. Kennett, J. Kurchan, P. Le Doussal, R. Monasson, M. 
J. Rozenberg  and G. Semerjian
for collaborations on related 
problems.
I acknowledge financial support from an Ecos-Sud travel grant, an ACI
project,
the J. S. Guggenheim Foundation and ICTP-Trieste, as well as
hospitality from the Universidad de Buenos Aires and Universidad
Nacional de La Plata, ICTP, and the KITP.

\end{document}